\documentclass[twocolumn]{article}
\usepackage[english]{babel} 
\usepackage[a4paper,top=2cm,bottom=2cm,left=2cm,right=2cm,marginparwidth=1.75cm]{geometry}
\usepackage[numbers]{natbib}
\usepackage{amsmath}
\usepackage{graphicx}
\usepackage[colorlinks=true, allcolors=black]{hyperref}
\usepackage{multirow}
\usepackage{wrapfig}
\usepackage{xspace}
\usepackage{array}
\usepackage{booktabs}
\usepackage{comment}
\usepackage{makecell}
\usepackage{paralist}
\usepackage{float}
\usepackage{listings, listings-rust}
\usepackage{cleveref}

\title{Targeted Fuzzing for Unsafe Rust Code: Leveraging Selective Instrumentation}

\author{
David Paaßen\\
david.paassen@uni-due.de\\
University of Duisburg-Essen\\
Germany
\and
Jens-Rene Giesen\\\
jens-rene.giesen@uni-due.de\\
University of Duisburg-Essen\\
Germany
\and
Lucas Davi\\
lucas.davi@uni-due.de\\
University of Duisburg-Essen\\
Germany
}

\date{}

\begin{document}

\newcommand{\toolname}{FourFuzz\xspace}
\newcommand{\toolnamelong}{Focus On Unsafe Rust Fuzzer\xspace}
\newcommand{\aflpp}{\texttt{AFL\nolinebreak\hspace{-.085em}\raisebox{.15ex}{\small +}\nolinebreak\hspace{-.10em}\raisebox{.15ex}{\small +}}\xspace}
\newcommand{\aflrs}{\texttt{afl.rs}\xspace}
\newcommand{\Cpp}{C\nolinebreak\hspace{-.085em}\raisebox{.2ex}{\small +}\nolinebreak\hspace{-.10em}\raisebox{.2ex}{\small +}\xspace}

\maketitle

\begin{abstract}
Rust is a promising programming language that focuses on concurrency, usability, and security. It is used in production code by major industry players and got recommended by government bodies. Rust provides strong security guarantees achieved by design utilizing the concepts of ownership and borrowing. However, Rust allows programmers to write unsafe code which is not subject to the strict Rust security policy. Empirical studies show that security issues in practice always involve code written in unsafe Rust.

In this paper, we present the first approach that utilizes selective code coverage feedback to focus the fuzzing efforts on unsafe Rust code. Our approach significantly improves the efficiency when fuzzing Rust programs and does not require additional computational resources while fuzz testing the target. To quantify the impact of partial code instrumentation, we implement our approach by extending the capabilities of the Rust compiler toolchain. We present an automated approach to detect unsafe and safe code components to decide which parts of the program a fuzzer should focus on when running a fuzzing campaign to find vulnerabilities in Rust programs. Our approach is fully compatible with existing fuzzing implementations and does not require complex manual work, thus retaining the existing high usability standard. Focusing on unsafe code, our implementation allows us to generate inputs that trigger more unsafe code locations with statistical significance and therefore is able to detect potential vulnerabilities in a shorter time span while imposing no performance overhead during fuzzing itself.

\end{abstract}

\section{Introduction}
\label{sec:introduction}
Software vulnerabilities such as memory safety issues are still one of the most common vulnerability types in modern software programs. They have been studied extensively over the last decades and have become a hot topic of research~\cite{SZE13}. 
Existing work either presents new defenses to prevent attacks or finds new and creative ways to circumvent existing implementations of protection techniques. Microsoft states that 70 \% of all security vulnerabilities are caused by memory safety issues~\cite{MSR2019}. Similar results have been reported by Google for the popular Chrome web browser~\cite{GOO2020}.
One way to address memory safety issues is to implement programs in memory-safe languages such as Rust, which is recommended by prominent industry players~\cite{CLA2022} as well as government bodies~\cite{CIS2023, WHO2024}. The Rust programming language is designed to check the code for potential memory unsafe code patterns such as null pointer dereferences at compile time. Alternatively, the Rust compiler adds the appropriate run time checks automatically to prevent exploitation of memory safety bugs. 
As most Rust security guarantees are checked during compile time, they do not impose any performance penalty when executing a program.
This has contributed to the rising popularity of Rust in production code, e.g., it is utilized by Mozilla~\cite{CLA2017}, Microsoft~\cite{WAN2023}, Cloudflare~\cite{GAL2019}, Dropbox~\cite{JAI2020}, Facebook~\cite{FAC2021}, and is also used to write Linux kernel modules which were introduced in version~6.1~\cite{KEE2022}.

However, the aforementioned safety guarantees cannot be upheld in all use-cases. For example, when a program requires dereferencing C-style raw pointers, arbitrary type casts, or needs to communicate with low-level systems~\cite{QIN2020}.
To allow for such code patterns, Rust includes the prominently marked \emph{unsafe} keyword, which disables certain security checks and enhances the capabilities of programmers. However, unsafe Rust no longer guarantees the safety of the program, and therefore may introduce issues well-known from C and \Cpp\ programs such as buffer overflows, use-after-frees, and undefined behavior in general, which may cause exploitable vulnerabilities. Thus, vulnerabilities caused by unsafe code are still an important issue in the Rust ecosystem. The Rust advisory database (RustSec)\footnote{\url{https://rustsec.org/}} tracks such vulnerabilities since 2016. It contains over 600 entries, of which 18 describe bugs that are located in the Rust standard library, which is thoroughly reviewed and commonly written by Rust experts.

Empirical studies~\cite{XU2021, QIN2020} which analyzed known vulnerabilities show that all memory safety issues found since the first stable release of the Rust compiler involve unsafe Rust code, and therefore conclude that safe Rust can be considered safe in practice. At first glance, it might be surprising that code regions explicitly marked as unsafe in Rust suffer from numerous vulnerabilities since one would assume that developers should be cautious in using the unsafe keyword, taking special care that the program cannot be exploited and ensuring that unsafe code regions are small to allow efficient and scalable inspection and review. Similarly, in the domain of trusted execution environments (e.g., Intel SGX), one would assume that security-critical code is less prone to safety issues.
However, recent research revealed that almost all public SGX enclaves suffer from security issues~\cite{BUL2019, CLO2020, KHA2020, CLO2022}.

\citet{QIN2020} systematically analyzed a set of 70 real-world memory safety vulnerabilities of Rust programs and found that the most prominent bug types are buffer overflows, null pointer dereferences, and use-after-free. According to the RustSec database, memory safety issues\footnote{Note that not all reports include a vulnerability type.} are by far the most common vulnerability type with 187 documented cases. The most common CVSS severity scores are high (134) and critical (72). This shows that security issues are still a significant problem in practice when deploying Rust programs that utilize unsafe code, even when this code is written by experienced Rust programmers.

\citet{AST2020} as well as \citet{EVA2020} analyzed the usage of unsafe code in Rust projects. They find that 24\% and 29\% of analyzed Rust programs contain unsafe code. Furthermore, around 50\% of Rust libraries depend on or utilize unsafe code. This number increases to 60\% when only considering the top 500 most popular crates from \emph{crates.io}. An unsafe code block is relatively small, as 75\% of unsafe blocks consist of at most 21 instructions. In general, unsafe code is used sparingly, i.e., 90\% of Rust crates have fewer than 2 unsafe functions and less than 11 unsafe code blocks. These results provide us with two key insights (1) Rust projects commonly rely on unsafe Rust code, and (2) most of the code of a Rust program is safe code and is therefore checked by Rust's security analysis.

It should be noted that it is not feasible to completely refrain from using unsafe Rust. Hence, programmers need other ways to ensure the security of their implementations, e.g., through extensive testing. One of the most popular techniques to detect bugs is fuzzing, i.e., randomly generated inputs are passed to a program, and a bug oracle checks for potentially unwanted behavior; usually a crash of the program under test~\cite{ZAL2019, BOE2016, FIO2020}.

Typically, the reasons for crashes are memory safety issues, e.g., because a code pointer is overwritten in case of a buffer overflow.
Fuzzers for C and \Cpp code commonly try to maximize the code coverage due to the fact that a fuzzer cannot find bugs in code that is never executed~\cite{ZAL2019, KLE2018}.
However, since Rust code is secure by default, this approach does not scale for Rust programs, where only small parts of a program are written in unsafe code blocks~\cite{XU2021,QIN2020}. Hence, by ignoring these safe code parts, we can optimize Rust code fuzzing to the few sections of a program that cause vulnerabilities and may lead to exploits such as remote code execution. This can be achieved using partial code instrumentation.

The most popular fuzzer for C and \Cpp programs is AFL~\cite{ZAL2019} which introduced the notion of coverage feedback to optimize the fuzzing process. Following the success of AFL, the authors of \aflpp~\cite{FIO2020} have since further optimized the design of AFL and implemented a plethora of improvements from the industry as well as academia. Due to the performance and capabilities of \aflpp, the Rust fuzzing team has since implemented \aflrs~\cite{ARS2016} which allows programmers to use \aflpp basic capabilities to fuzz Rust programs. However, even though \aflrs is based on \aflpp it does not support the whole feature set of \aflpp. Most notably, it is not possible to use the highly optimized and feature-rich \aflpp compiler suite as these are only compatible with C and \Cpp code but not Rust. This includes the capabilities of limiting the coverage feedback instrumentation to certain parts of the program.

To the best of our knowledge, the only existing academic fuzzer for Rust code has been implemented by \citet{CRU2023} as part of a registered report. The fuzzer is called \textsc{CrabSandwich} and is not tailored to focus on unsafe Rust code. Thus, it does not prioritize Rust code that potentially causes security issues. Furthermore, \textsc{CrabSandwich} is currently only a preliminary implementation and is not publicly available. Other existing publications regarding Rust security use static analysis based on different intermediate code representations~\cite{BAE2021, ZHU2021}. However, static analysis tools do not provide proof-of-vulnerabilities and suffer from many false positive detections. Thus, they require significant effort to detect exploitable vulnerabilities.

In this paper, we present and implement \toolname (\toolnamelong), the first fuzzer that is specifically tailored to prioritize unsafe Rust code. We implement partial instrumentation to focus the fuzzing efforts on unsafe Rust code to significantly increase the efficiency and probability of triggering and detecting vulnerabilities in Rust programs.

To implement partial code instrumentation in Rust we need to patch the Rust compiler itself.
This is challenging as Rust uses a complex compilation toolchain:
This includes multiple intermediate representations and several compiler stages facilitating Rust-specific code analysis.
Further, the Rust toolchain leverages the LLVM compiler suite to generate machine code.
In total, the Rust toolchain comprises well over 10 million lines of code written in various programming languages.

Furthermore, our patched Rust compiler automatically exports the list of functions that contain unsafe code, which we subsequently use as an additional input in the instrumentation phase. This helps the fuzzer to focus on inputs that may execute code that leads to potential vulnerabilities instead of mutating all inputs that achieve new coverage, even in code parts that exclusively contain safe Rust code. To implement this approach, we only require call graph data which enables us to utilize \toolname even on large and complex Rust programs.

In our evaluation, we compare our implementation against the existing implementation of \aflrs\ on a test set of 10
programs using a run time of 24 hours,
and performing 30 repetitions.
During our evaluation, we follow existing best practices and use statistical significance tests as well as standardized effect sizes to assess a fuzzer's performance.
We also quantify the impact of partial instrumentation when fuzzing unsafe Rust code.
Due to the fact that \aflrs and \toolname share the same code base, we can ensure that the performance differences are caused by the fuzzer's ability to focus on unsafe code.
We find that \toolname outperforms \aflrs\ on eight
targets with statistical significance by executing unsafe code parts more often while requiring a shorter time span to do so. 
Our evaluation further shows that, on average, \toolname only requires instrumentation of around 20\%\
of the program functions to collect accurate coverage information for paths that lead to unsafe code locations. Furthermore, \toolname is able to generate inputs that trigger 15\%\
more unsafe code locations on average. Thus we significantly improve the overall performance of Rust fuzzing. Note that we achieve these performance gains without any computational overhead during the fuzzing process itself, as \toolname does not require any additional run time code to work properly. 

Furthermore, we present the first study to systematically evaluate and quantify the impact of partial instrumentation on real-world Rust software. Our results show that partial instrumentation has an acceptable build time overhead (even on complex targets) while significantly improving the overall performance when analyzing the ability of a fuzzer to trigger code written in unsafe Rust.

\section{Background}
\label{sec:backgroud}
First, we explain the fundamentals of the Rust programming language and how it can provide broad security guarantees while at the same time being as flexible as C and \Cpp. We further introduce the concept of unsafe Rust code and the notion of fuzzing and how it has been used in existing scientific studies.

\subsection{Rust}

Rust is a powerful systems programming language that emphasizes on performance while providing strong safety guarantees. Contrary to other memory safe languages, such as Java, Rust allows the programmer to precisely control memory allocations and is versatile enough to be used as a low-level systems programming language. 

Rust achieves memory safety through its combination of a strict type system and the ownership/borrowing model.
This improves the performance of Rust compared to other memory safe languages that require, e.g., a garbage collector.
Rust originates from Mozilla Research but is currently maintained by the Rust Foundation which is a non-profit organization founded by tech companies such as Amazon, Google, Huawei, and Microsoft. All of which use Rust as part of their software stack. For example, Google supports Rust to write native OS components for Android~\cite{AND2022}.
Similarly, Microsoft supports Rust to write Azure applications~\cite{MIC2023} as well as the implementation of driver software for Windows~\cite{WAN2023} citing Rust's security features as a major reason for using it.
Furthermore, Rust code has also been included as part of the Linux kernel since version 6.1 which allows contributors to write components such as kernel modules in Rust~\cite{KEE2022}.

\subsubsection{Unsafe Rust}
\label{subsubsec:unsafe_rust}

At its core, Rust achieves temporal memory safety by restricting aliasing of mutable data. This prevents memory safety issues such as data races, use-after-free, and double-free. During compilation Rust guarantees that only a single \emph{mutable} reference exists for each variable at any point in the program's execution. The compiler does not limit the number of read-only references as long as \emph{no} mutable references exist at the same time in the same context.

Spatial memory safety is achieved via a combination of compile time and run time checks. The Rust compiler determines the bounds of any object in memory and verifies that memory accesses are in bounds of the respective object. If the size of the memory object is unknown at compile time, the Rust compiler automatically inserts the appropriate bounds checks. Any access beyond an object's allocated memory results in a panic.

The design of safe Rust severely limits the capabilities of a programmer, for example, when data needs to be shared with other (unsafe) programs/libraries or when implementing certain code patterns, e.g., a mutex.
Therefore, Rust supports a keyword called \texttt{unsafe} which lets programmers ignore certain safety restrictions, e.g., modify raw pointers or call unsafe functions. 
However, even if code is marked as \texttt{unsafe}, Rust still checks the respective code but cannot guarantee its memory safety. 
Note that not all functions that contain \texttt{unsafe} code also have to be marked as \texttt{unsafe}.
Instead, Rust requires the programmer to ensure that all public facing functions do not induce unwanted behavior when the correct types are used (which can be easily checked by the compiler).
This code pattern is called a \emph{safe abstraction} and allows programmers to call functions that use unsafe code without tagging such calls as unsafe~\cite{AST2020}. In practice, unsafe Rust code is commonly used to communicate with foreign code, the kernel, hardware components, or to improve the codes' performance as well as optimizing memory management~\cite{FUL2021}. Due to the safe abstraction design pattern, a programmer may use unsafe code unknowingly (see Section~\ref{subsubsec:design:detect_unsafe:example}).

\subsubsection{Rust Compilation Pipeline}

\begin{figure*}[htb]
\center
\includegraphics[width=0.9\textwidth]{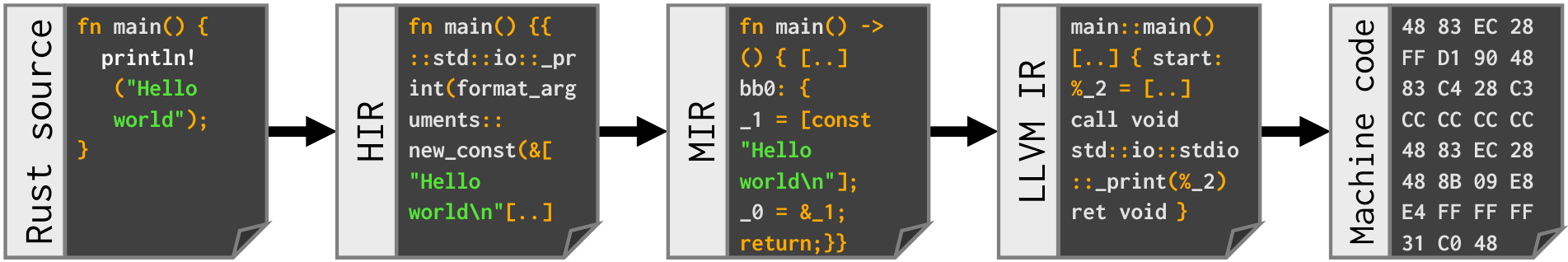}
\caption{Simplified overview of the Rust compilation pipeline.}
\label{fig:rustcomp}
\end{figure*}

The Rust compilation process includes multiple complex stages to translate Rust code to highly optimized machine code (see Figure~\ref{fig:rustcomp}) that can be securely executed on a large set of CPU architectures and platforms. Rust ensures type safety using the high-level intermediate representation (HIR) while the borrow checker is executed on the mid-level intermediate representation (MIR) which is basically a CFG representation of the code. After transforming the code into the LLVM IR, Rust utilizes the LLVM compiler suite to optimize the code which is subsequently transformed into the final binary representation.

\subsection{Fuzzing}

Fuzzing has become one of the most popular testing methods, especially for C and \Cpp programs. In this section, we discuss the most important aspects of fuzzing in the context of testing Rust code.

Fuzzing has been popularized in recent years mostly by the success of AFL, a mutation-based gray-box fuzzer. Besides gray-box fuzzing, there also exists white-box and black-box fuzzing. A black-box fuzzer simply generates inputs without any knowledge of the program internals or the program source code. This means that a black-box fuzzer (e.g., zzuf~\cite{HOV2006}) has a very high throughput but cannot reason about the quality of the generated input apart from program crashes. White-box fuzzers (e.g., SAGE~\cite{GOD08}) use extensive program analysis to be able to generate high quality inputs that, for example, solve certain path constraints and therefore increase code coverage. This usually means that the throughput is rather low, and commonly requires access to the program's source code. AFL~\cite{ZAL2019} and other gray-box fuzzers try to find a middle ground between white-box and black-box fuzzing by utilizing simple code analysis techniques (typically coverage information) that do not require expensive computation while fuzzing. Over the years, scientists proposed various different fuzzers~\cite{ASH2019, MAN2021, SCH2021, FIO2022, ZHU2022}, to improve the general performance and explore different application domains where fuzzing can be used~\cite{MAN2021, ZHU2022}.

Existing best practices~\cite{APP2024} recommend the usage of in-memory fuzzing which significantly improves the execution speed as it removes the need to fork the target program for every execution. However, to use in-memory fuzzing the fuzzer requires a so called fuzzing harness which in its most basic form calls a test function (e.g., data processing function) with the input bytes which are generated by the fuzzer. The authors of \aflrs only support the usage of in-memory fuzzing. Thus, to test a Rust program, one always needs to write a fuzzing harness (also called fuzz target) which is compiled into a specially instrumented binary.

\section{\toolname}
\label{sec:design}
\begin{figure}[htb]
\center
\includegraphics[width=0.95\columnwidth]{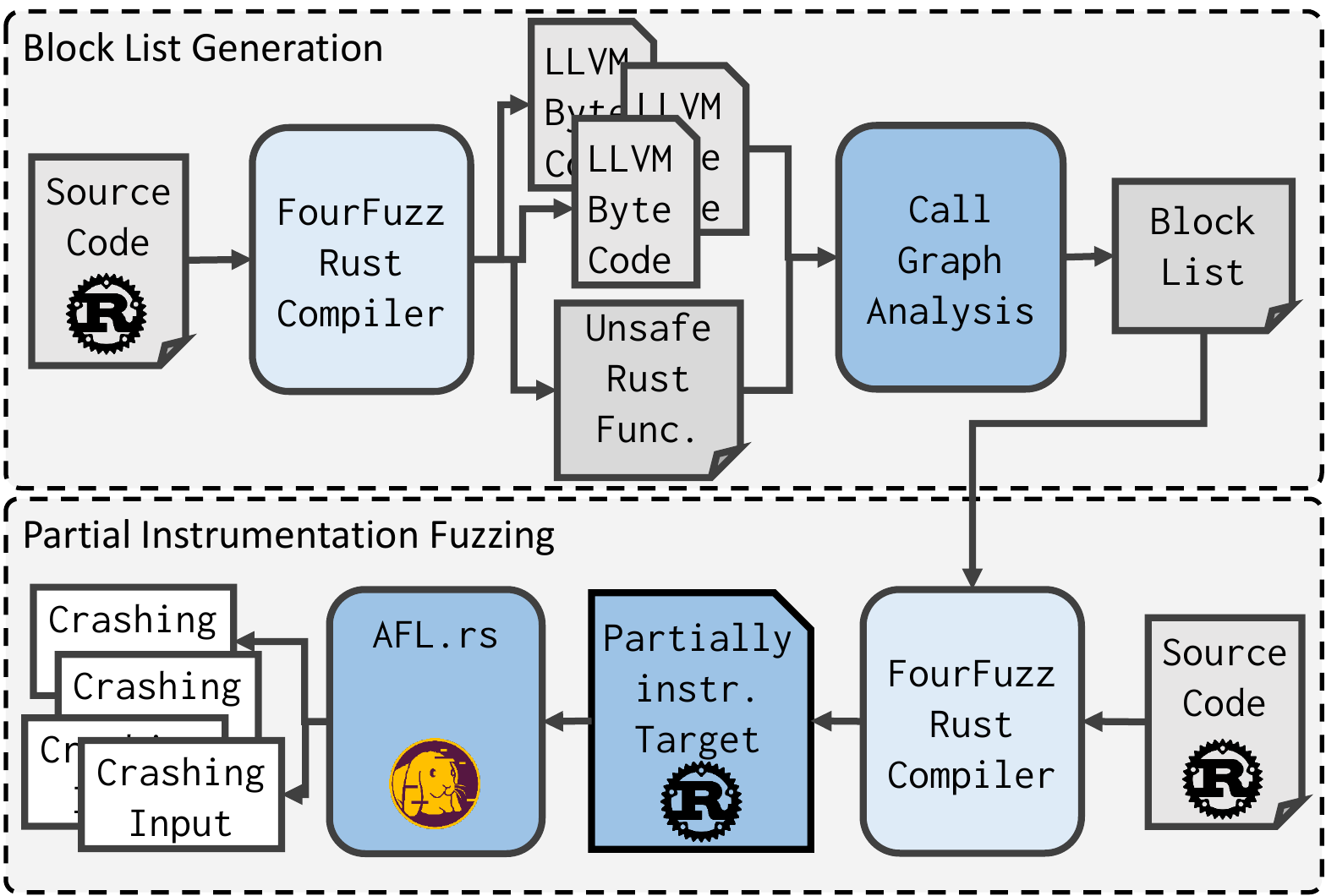}
\caption{Workflow of \toolname. First, we generate a block list which is subsequently used to build and fuzz a partially instrumented target.}
\label{fig:overview}
\end{figure}

Generally, the concept of selective code instrumentation can be implemented for any fuzzer that utilizes code coverage feedback. According to the Rust fuzzing authority~\cite{RUS2024}, most of real-world bugs detected in Rust programs have been uncovered by \aflpp~\cite{FIO2020}, \texttt{libfuzzer}~\cite{LLV2018}, and \texttt{honggfuzz}~\cite{GOO2024} for all of which exists a Rust compatible implementation, namely \aflrs~\cite{ARS2016}, \texttt{cargo-fuzz}~\cite{CAR2024}, and \texttt{hongfuzz.rs}~\cite{HON2024}. We opted to use \aflrs for our implementation \toolname, due to the fact that the underlying fuzzer (i.e., \aflpp) is well maintained and provides the most comprehensive feature set which allows one to utilize a large number of fuzzing optimizations published in academic publications. In the following we describe the different components required to run \toolname. We explain how \aflrs works internally and how we deal with the challenges of implementing partial instrumentation and the detection of unsafe Rust Code itself. Subsequently, we describe the overall design of \toolname. The high level workflow of \toolname is depicted in Figure~\ref{fig:overview}.

\subsection{Afl.rs}

The main purpose of \aflrs is to invoke the Rust compiler with the correct flags to build the program under test with the correct instrumentation that allows \aflpp to retrieve coverage information when fuzzing the target binary. Additionally, \aflrs provides an easy-to-use wrapper code that allows to invoke \aflpp within the Rust toolchain. Using its default configuration, \aflrs (1) enables additional checks (e.g., overflow checks) to improve the bug finding capabilities (2) enables code optimization up to level three to improve the run time performance, and (3) sets LLVM flags to enable code coverage feedback. Recently, \aflrs added the LLVM trace option for compare instructions, which implements a CmpLog-style instrumentation originally implemented in Redqueen~\cite{ASH2019}. Lastly, \aflrs adds the AFL LLVM run time library to the target binary which is responsible to run initialization code, to communicate with \aflpp, and provide functionality required by the instrumentation.

\subsection{Partial Instrumentation}

\aflpp allows the programmer to utilize partial code instrumentation in two different ways. First, via the \texttt{\_AFL\_\allowbreak COVERAGE}\allowbreak-function family which allows a programmer to selectively enable the instrumentation for certain parts of the program by tagging these directly in the source code, i.e., by calling \texttt{\_\_AFL\_\allowbreak COVERAGE\_\allowbreak ON()} and \texttt{\_\_AFL\_COVERAGE\_OFF()} functions, respectively. However, \aflrs uses LLVM-based code coverage, which does not support calling these AFL specific functions, therefore one cannot utilize them to compile Rust code. Secondly, \aflpp allows to directly whitelist and blacklist parts of the code using \texttt{AFL\_LLVM\_ALLOWLIST} and \texttt{AFL\_LLVM\_DENYLIST}, which are text files containing a list of the code parts that should be instrumented or ignored. These lists are processed when compiling a target program using \texttt{afl-clang-fast} or \texttt{afl-clang-lto}.
Even though Rust, like \aflpp, utilizes LLVM in its compilation toolchain it does not support the direct usage of any specialized \aflpp compiler and instead only supports the use of LLVM instrumentation.

The LLVM compiler itself also provides two different partial instrumentation features. The first method requires editing the source code, namely adding a special \texttt{no\_sanitize} attribute to the definition of all functions that should not be instrumented. This attribute however, is a \texttt{clang} feature that is not available to the Rust compiler. The second way is a blacklist and whitelist feature that does not require source code changes and is utilized via special flags that either specify the usage of an \texttt{allowlist} or \texttt{ignorelist}. Adding these flags anywhere in the compilation process of a Rust program triggers an exception, as they are not supported by the LLVM binaries used by the Rust compilation toolchain.

Thus, it is not possible to use any of the existing partial instrumentation techniques currently implemented in \aflpp\ or LLVM with programs written in Rust. Instead, we have to implement selective instrumentation capabilities directly into Rust. As a block list is more suitable for an automated approach like ours, we decide to implement it instead of a source code based method. 
\aflpp supports various instrumentation techniques.
However, the only variant that is supported by \aflrs is called \texttt{sanitizer-coverage-trace-pc-\allowbreak guard}, and is a feature of LLVM which adds a call to a coverage feedback function at every CFG edge. This feedback function reads a dynamically generated edge-id, which is used as an index into the coverage map of the respective program. The coverage map is evaluated by \aflpp after each fuzz run to decide if a mutated input increased the coverage of the program code.

As the coverage feedback utilized by \aflrs is a LLVM feature, we implement the block list inside Rust's LLVM module, mainly inside the \texttt{SanitizerCoverage} component.
This feature adds the previously mentioned calls to the coverage feedback function, which in turn sets the respective entry in the shared coverage map.
Similar to \aflpp and LLVM itself, our implementation reads a text file that contains a list of functions that should not be instrumented.
In our particular use case, this is a list of functions that do not lead to any unsafe Rust code.
However, our implementation can also be used for other use cases, e.g., to improve patch testing via fuzzing.
Additionally, our implementation can be used by any fuzzer that supports LLVMs \texttt{sanitizer-coverage-trace-pc-guard} instrumentation and similar instrumentation features (e.g., basic CmpLog-style instrumentation) as the partial instrumentation feature is implemented in a LLVM component.
Due to the fact that we implement our design as part of Rust and LLVM, respectively, \toolname compiles its own Rust compiler suite which it subsequently utilizes to generate the partially instrumented target binary.

\subsection{Detecting Unsafe Rust Code}
\label{subsubsec:design:detect_unsafe:example}
Intuitively, one might get the impression that searching the source code of a Rust project for the \emph{unsafe} keyword is sufficient to assess whether the project contains unsafe Rust code.
However, this approach only detects unsafe code that exists in the exact Rust projects' source code, but it cannot check the dependencies a project relies on.
Thus, searching only inside the Rust project completely ignores the fact that Rust crates, i.e., dependencies, commonly contain unsafe Rust code.
The Rust build tool \texttt{cargo} manages all crates and automatically downloads their transitive dependencies during compilation. 
Rust supports a concept called \emph{safe abstraction} (cf. \cref{subsubsec:unsafe_rust}), which allows developers of Rust crates to wrap potentially unsafe operations in safe interfaces.
Commonly, the safe interfaces are ordinary Rust functions, i.e., a developer cannot decide whether a function is just a function or a safe abstraction for unsafe code. 
As a result, a developer may unknowingly include unsafe code in a program.
For example, a call to any third party function may execute unsafe code.
In practice, \citet{EVA2020} find that around 50\% of all Rust libraries contain unsafe code, and only 29\% of all analyzed Rust libraries use unsafe code directly.
Thus, the likelihood of unknowingly calling unsafe code is high.

\subsubsection{Motivating Example}

To illustrate the problem of unknowingly calling unsafe Rust code, consider our example in \Cref{lst:unsafe_caller}.
At \Cref{line:call_crate}, the program calls a function which it imports from the \texttt{byteops} crate, using parameters that a user provides in \Cref{line:index,line:value}.
In our example, this code snippet represents the \emph{entire} code base of a Rust project. 
As it does not contain the \texttt{unsafe} keyword, it is not evident to the reader of \Cref{lst:unsafe_caller} that this project relies on unsafe code.
\Cref{lst:unsafe_crate} shows the source code sample of our \texttt{byteops} crate.
The code defines the \texttt{Bytes} data structure that operates on arrays of bytes and stores a start pointer, an end pointer, and a cursor at \Cref{line:struct_def}.
The \texttt{store\_at} function calculates the value of a pointer \texttt{ptr} and uses it to store an arbitrary value that is provided as a function argument. 
Both operations require unsafe code which is hidden behind the seemingly safe abstraction. However, \texttt{store\_at} does not perform any bounds checking, thus an attacker might be able to write arbitrary values to arbitrary memory, even though the code base in \Cref{lst:unsafe_caller} is exclusively written in safe Rust which prevents such vulnerabilities.

\begin{figure}[tbp]
\begin{lstlisting}[caption={Example program that unknowingly calls unsafe Rust code.}, label=lst:unsafe_caller]
use std::io;
@\label{line:use_crate}@use byteops::Bytes;
@{\scriptsize{[...]}}@
fn input() -> u64 {
    let mut input = String::new();
    io::stdin().read_line(&mut input).expect("@\scriptsize{...}@");
    input.trim().parse().expect("@\scriptsize{...}@")
}

fn main() {
    let v: Vec<u8> = vec![1, 2, 3];
    let mut b = Bytes::new(&v);
@\label{line:index}@    let index = input() as usize;
@\label{line:value}@    let value = input() as u8;
@\label{line:call_crate}@    b.store_at(index, value);
}
\end{lstlisting}
\end{figure}
\begin{figure}[bp]
\begin{lstlisting}[caption={Example code of our \texttt{byteops} crate that includes unsafe Rust.}, label=lst:unsafe_crate]
pub struct Bytes {@\label{line:struct_def}@ start: *const u8, 
  end: *const u8, cursor: *const u8,
}

impl Bytes {
    pub fn new(slice: &[u8]) -> Bytes {@\scriptsize{...}@}
    @\label{line:fn_def}@fn store_at(&self, n: usize, v: u8) {
        unsafe {
            @\label{line:ptr_add}@let ptr = self.cursor.add(n);
            @\label{line:ptr_write}@std::ptr::write(ptr, v);
        }
    }
    @{\scriptsize{[...]}}@
}
\end{lstlisting}
\end{figure}

\subsubsection{Rust Compiler Extension}

To ensure that \toolname can detect all unsafe code, we need to extend the capabilities of the Rust compiler to generate a list of \emph{all} functions that contain unsafe Rust code.
As mentioned before, the Rust compilation pipeline translates the source code into different intermediate representations, namely HIR, MIR, and LLVM IR.
These representations mainly differ in their capabilities regarding static program analysis.
First, we need to decide where inside the complex Rust compilation process we should add our analysis pass which can be applied at HIR, MIR, and LLVM-IR level.
The HIR stores the code before any code optimizations happen.
This impedes the identification of functions further down the compilation process, as we need to identify the functions using low-level symbol names which are used at the LLVM instrumentation stage.  However, Rust's function names are changed after running the HIR stage and therefore extracted function names might be different at the point the block list is utilized.
We cannot directly operate on the LLVM IR because at this stage the compiler removed any meta information about the unsafe status of instructions as LLVM does not support this code property\footnote{As such a property is not required in other languages supported by LLVM such as C, \Cpp, or Swift.}.
Since the majority of Rust specific optimizations happen on MIR-level, and meta information about the unsafe status of functions and statements is still available, we implement our design at this stage.

To implement our analysis, we extend the \texttt{rustc\_monomorphize} crate, specifically the \texttt{collect\_and\_partition\_mono\_items} function. 
The Rust compiler calls this function before code generation. We collect all function instances, including specialized instances of generic functions, and analyze the basic blocks of each instance to check for unsafe code at the instruction-level. Due to the fact that the terminator is not considered part of a basic block, we check it separately.
Note that our implementation is not affected by potential false negative detections of unsafe instructions~\cite{CHO2024} because our analysis does not rely on the safety status of basic block objects, but on the safety of each individual instruction.

Once we detect at least one instruction that is unsafe, we retrieve and store the symbol name of the surrounding function. 
\toolname further processes this file during its path finding process to generate the block list that is subsequently used in the partial instrumentation phase. Note that programmers may still want to ignore (unsafe) dependency code during fuzz testing (e.g., because dependency code is tested separately) which is also supported by \toolname.

\subsection{Design}

\begin{figure}[htb]
\center
\includegraphics[width=0.95\columnwidth]{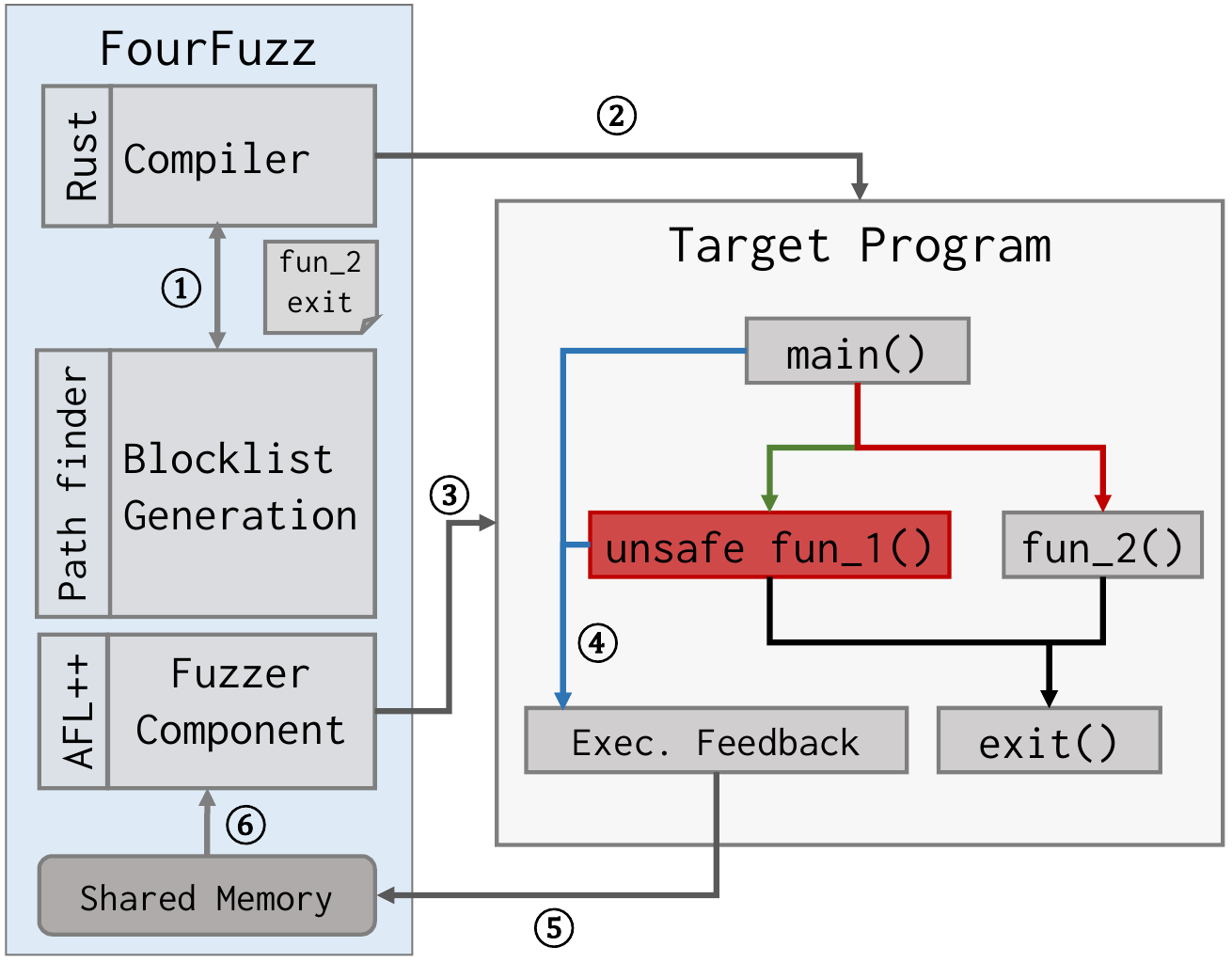}
\caption{Overview of the design of \toolname and its partial instrumentation approach.}
\label{fig:aflrsov}
\end{figure}

In the following, we describe the different modules of \toolname and how they interact with one another.  
An overview of how our design and its components work are depicted in Figure~\ref{fig:aflrsov}. 
In the first step, our modified Rust toolchain compiles the target program to generate the call graph (CG) based on LLVM binary code as well as a list of functions that contain unsafe code. To generate a complete call graph, \emph{path finder} merges the call graphs of different program modules utilizing the fact that Rust generates unique names for each function, including a hash based on the functions content. This ensures that \toolname does not miss any paths inside the compiled program that leads to unsafe Rust code. The \emph{path finder} component leverages the CG and the list of unsafe functions to construct our block list, i.e., the set of functions that will \emph{never} reach any \emph{unsafe} Rust code. Subsequently, \emph{path finder} traverses the whole-program CG (i.e., every function) and matches the CGs' nodes with the list of functions that contain unsafe code. If no paths from the function to unsafe code exist, the respective function is added to the block list.
This block list is subsequently used by the Rust compiler (more specifically the LLVM instrumentation pass) to only add coverage feedback calls to functions that actually reach unsafe code.
Note that for the example in Figure~\ref{fig:aflrsov}, \toolname will ignore inputs that execute \texttt{fun\_2} as this function does not contain or lead to any unsafe Rust code.
This allows \toolname to focus on a smaller set of inputs and thus create more inputs that potentially trigger a vulnerability inside \texttt{fun\_1}.

Note that \toolname works as a drop-in replacement for the Rust compiler as well as \aflrs, i.e., a user can use the exact same commands to compile and fuzz the target binary. Therefore, we argue that \toolname does not impose additional significant usability issues, reported in other fuzzer implementations~\cite{LI2021}.

\toolname's approach to partial instrumentation requires compiling the target twice in succession.
However, Rust builds are not reproducible in all use cases and depend on multiple factors, for example, the absolute build path.
In fact, issues with reproducibility are wide-spread enough for the Rust developers to maintain a separate label in the issue tracker for issues related to reproducibility\footnote{\href{https://github.com/rust-lang/rust/labels/A-reproducibility}{https://github.com/rust-lang/rust/labels/A-reproducibility}}.
To prevent issues of reproducibility from diminishing the effectiveness of \toolname, we verify that builds of our fuzzing targets are compatible with each other. e.g., between the execution of \emph{path finder} and the partial instrumentation step.

\subsubsection{Implementation}

Our implementation works on the Rust stable release version 1.77 which uses LLVM version 17 to generate the binary code. To fuzz the target programs, \toolname utilizes \aflrs in version 0.15.3 which is based on \aflpp version 4.10c. We implemented our changes to the Rust compiler as part of the MIR code passes, a modified LLVM code pass, as well as a Python script that implements the path finder component as well as a custom Python framework to run the experiments.

\section{Evaluation}
\label{sec:evaluation}
Evaluating fuzzers is a non-trivial problem~\cite{KLE2018, BOE2020} as the majority of fuzzing research does not follow existing recommendations~\cite{SCH2024}. When evaluating the impact of partial code coverage instrumentation, we want to ensure that our results are scientifically sound.
Hence, we follow existing best practices~\cite{KLE2018} and use a timeout of 24h 
and repeat every experiment 30
times. We further utilize the recommended statistical significance test (Mann–Whitney U test) to ensure that differences measured in our experiments are not the result of randomness. Additionally, we use standardized effect sizes (Vargha and Delaneys $\hat{A}_{12}$ statistic~\cite{ARC2014}) to address the fact that the programs in our test set feature different amounts of unsafe code locations (see Table~\ref{tab:res1}). As our fuzzer prioritizes functions that contain unsafe Rust code, we utilize the time it takes a fuzzer to find such a location as an evaluation metric, analogous to the time a fuzzer requires to trigger a bug. Thus, the goal of our evaluation is to show that \toolname generates inputs that execute more locations that contain unsafe Rust code in a shorter time span compared to the existing state-of-the-art.

\begin{table}[tbp]
\center
\caption{Overview of our test set, including a short description, and number of downloads on the main Rust library registry \emph{crates.io}.}
\footnotesize
\begin{tabular}{lp{4cm}r}
\toprule
Project & Description & Downloads \\ 
\midrule
capnproto & Data serialization framework. & 3.7 mil. \\
httparse & HTTP 1.x protocol parser. & 152 mil. \\
image & Basic image processing library. & 31 mil. \\
lz4\_flex & Compression algorithm. & 15 mil.\\
rust-cssparser & Rust CSS Syntax Module. & 9 mil. \\
quiche & QUIC and HTTP/3 impl. by Cloudflare. & 0.33 mil. \\
rocket & Asynchronous web framework. & 5 mil. \\
syn & Generation of Rust Syntax Trees. & 530 mil. \\
toml & TOML decoder and encoder. & 188 mil. \\
ruzstd & Decoder for zstd compression. & 6.5 mil. \\
\bottomrule 
\end{tabular}
\label{tab:testset3}
\end{table}

To conduct our evaluation, we require a diverse set of programs that contain unsafe Rust code. Our target set consists of 10
popular Rust projects. Note that this is above the average of 8.9 targets reported by \citet{SCH2024} who did a literature review of 150 fuzzing papers published at renowned academic conferences. The test set incorporates a wide variety of different programs ranging from a library to handle HTML requests to code used for data compression. We provide an overview of our test set in Table~\ref{tab:testset3}. 

For each program in our test set, we collect a seed set of 100 different inputs which \aflpp and \toolname use as an initial set for mutations. 
We selected programs based on the following criteria:
\noindent
\begin{inparaenum}[(1)]
	\item The program contains at least one instance of unsafe Rust code that is not trivial to reach, i.e., this excludes targets that execute all unsafe Rust code regardless of the provided (valid) input.
	\item The program has to be either popular in the Rust community, i.e., over 1 million downloads on \emph{crates.io}, or it has to be implemented by a major industry player. This makes sure that our results are relevant in practice.
	\item The program is not a test or otherwise experimental software. This ensures that our results are relevant to real-world applications.
	\item The program is purely written in Rust, as we consider C and \Cpp code to be out of scope for our Rust fuzzer. Note that C/\Cpp functions can be fuzzed separately by a wide selection of existing fuzzers including \aflpp~\cite{FIO2020} and LibAFL~\cite{FIO2022}.\footnote{Note that mixed code binaries can further be protected by existing memory isolation approaches~\cite{LAM2017, LIU2020, RIV2021, BAN2023} that separate Rust and C/C++ memory which prevents malicious access from the C/C++ to the Rust code.}
\end{inparaenum}

We provide data and additional information related to our experiments in a dedicated code repository available at \url{https://github.com/uni-due-syssec/target-unsafe-rust}.

\subsection{Partial Instrumentation}

\begin{table}[htb]
\center
\caption{Number of functions in each project and the number of functions that do not lead to the execution of unsafe code.}
\vspace{10pt}
\footnotesize
\begin{tabular}{lrrr}
\toprule 
Project & Functions & \makecell{Excluded\\functions} \\ 
\midrule 
capnproto & 1752 & 1041 (59.42\%) \\
httparse & 380 & 356 (93.68\%) \\
image & 3689 & 3558 (96.74\%) \\
lz4 & 2731 & 2719 (99.56\%) \\
quiche & 3301 & 3132 (94.85\%) \\
rocket & 19322 & 19185 (99.29\%) \\
rust-cssparser & 5343 & 5282 (98.85\%) \\
syn & 2055 & 782 (38.05\%) \\
toml & 1089 & 1066 (97.89\%) \\
ruzstd & 2974 & 916 (30.80\%) \\
\bottomrule 
\end{tabular}
\label{tab:testset2}
\end{table}

First, we evaluate the partial instrumentation approach to fuzz Rust programs by comparing the common instrumentation using the default LLVM code coverage feedback with the selective instrumentation of \toolname that excludes all functions that cannot reach unsafe code parts.
The results are depicted in Table~\ref{tab:testset2}. 
Note that the results are based on the functions available in the binary code (i.e., optimized LLVM bytecode) and not on the number of functions in the Rust source code. We observe that, on average, the block list of \toolname contains around 81\%\
of all program functions, i.e., most functions of a Rust program will never reach any unsafe code and therefore do not need to be instrumented when fuzzing to find memory safety issues. This demonstrates the effectiveness of \toolname as the number of instrumented functions can be significantly reduced when fuzzing exclusively unsafe code.

\subsection{Reaching unsafe code}

The most common evaluation metrics when testing fuzzers either try to show that a fuzzer is able to trigger more bugs in a shorter time span or compare the code coverage achieved over the course of the experiment using different code coverage metrics~\cite{KLE2018}. However, the goal of \toolname is to improve the efficiency when fuzzing Rust programs, namely to increase the number of unsafe locations the fuzzer can trigger and decrease the time span it takes to generate such an input. Hence, we instrument each unsafe code location with an execution oracle to be able to detect when an input executes the respective unsafe code block or function for the first time. Instead of the time it takes a fuzzer to trigger a bug, we measure the \emph{time} it takes to find an input which executes a unsafe code location.

To assure a fair comparison, we follow existing recommendations~\cite{PAA2021, SCH2024} and use the fuzzer that \toolname is based upon (namely \aflrs) as a baseline which allows us to accurately attribute performance differences to our design choices.
We run all our experiments using Ubuntu 22.04 LTS on a server with an Intel Xeon Gold 6326 CPU with 32 cores and 256 GB of memory. We utilize Docker to ensure a fair and equal testing environment, which allows us to precisely allocate the same resources for each experiment.

We use statistical significance tests to assess if the results we measured can be attributed to the difference in the design or may be caused by randomness alone. 
As recommended by \citet{KLE2018}, we use a p~threshold of $0.05$ to determine if a result is statistically significant or not. 
As effect size threshold we follow the recommendations of \citet{VAG2000}.

\begin{table*}[htbp]
\center
\caption{Total number of non-trivial unsafe locations in each program and corresponding number of times \toolname outperforms \aflrs\ and vice versa. Additionally, the number of times the corresponding effect size is considered small, medium, or large. The table also shows the average $\hat{A}_{12}$ effect size for all statistical significant results.}
\begin{tabular}{lcccccccc}
\toprule 
\multirow{2}{*}{Project} & \multirow{2}{40pt}{\centering \#unsafe locations} & \multicolumn{2}{c}{\#stat. sig. results} && \multicolumn{3}{c}{Effect Size} & \multirow{2}{*}{Avg. $\hat{A}_{12}$} \\
\cline{3-4} \cline{6-8}
 & & \aflrs & \toolname && small & medium & large &  \\
\midrule 
capnproto & 7 & 0 & 7 && 0 & 0 & 7 & 0.99 \\
httparse & 14 & 0 & 0 && 0 & 0 & 0 & 0.00 \\
image & 2 & 0 & 2 && 0 & 0 & 2 & 0.88 \\
lz4\_flex & 2 & 0 & 2 && 0 & 0 & 2 & 1.00 \\
rust-cssparser & 4 & 0 & 4 && 0 & 0 & 4 & 0.99 \\
quiche & 11 & 0 & 10 && 0 & 0 & 10 & 0.75 \\
rocket & 2 & 0 & 1 && 0 & 1 & 0 & 0.69 \\
syn & 3 & 0 & 0 && 0 & 0 & 0 & 0.00 \\
toml & 14 & 0 & 12 && 0 & 0 & 12 & 0.76 \\
ruzstd & 1 & 0 & 1 && 0 & 1 & 0 & 0.66 \\
\bottomrule
\end{tabular} 
\label{tab:res1}
\end{table*}

We provide an overview of the experimental results considering the time it took each fuzzer to execute the unsafe code locations in Table~\ref{tab:res1}. We notice that \toolname has a statistically significant better performance for at least one unsafe location on eight of the ten test programs. For example, when comparing the fuzzer performance for \texttt{capnproto}, \toolname requires less time to generate inputs that execute all seven unsafe code locations with statistical significance for 99\% of all fuzz runs. Notably, \toolname never performs worse on any of the unsafe code locations with statistical significance. Looking at the effect sizes, we find that for most (37) unsafe locations the performance difference is considered large. The evaluation confirms that \toolname performs significantly better when generating inputs that trigger unsafe Rust code compared to \aflrs on the vast majority of targets. As \toolname and \aflrs use the same code base we conclude that the performance differences are caused by partial instrumentation helping \toolname to focus its resources on unsafe code locations.

\begin{figure}[htb]
\center
\includegraphics[width=0.95\columnwidth]{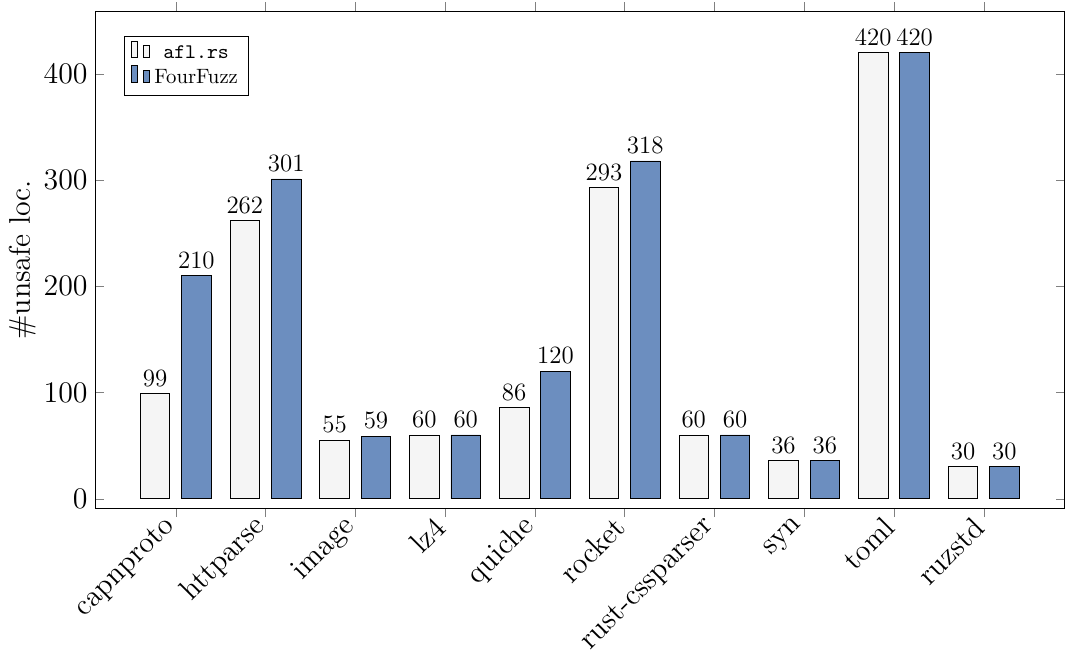}
\caption{Total number of times each fuzzer generates an input that executes a coverage oracle, i.e., unsafe code in the respective target project.}
\label{fig:unsafe_loc_trig}
\end{figure}

Additionally, we evaluate the total number of times each fuzzer successfully generates an input that reaches an unsafe code location over all runs. This helps us to better assess the performance of partial instrumentation by comparing the ability of each fuzzer to cover unsafe code locations at all. The results are depicted in Figure~\ref{fig:unsafe_loc_trig}. We observe that \toolname is able to trigger more unsafe code oracles over the course of our experiments on five of the ten targets. For the other targets, both fuzzers perform equally well over the course of 24h. Again we take a closer look at the results for \texttt{capnproto} which contains a total of seven unsafe code locations in the tested binary. On each of the 30 trials, a fuzzer can generate an input to execute each unsafe code location. We find that \toolname is able to trigger all unsafe code locations on all 30 trials (total of 210 unsafe code oracle hits) while \aflrs is unable to execute a total of 111 unsafe code locations. Overall, \toolname is able to detect 15\% 
more unsafe code locations on our test set compared to \aflrs. This demonstrates that partial instrumentation not only improves the time it takes a fuzzer to execute unsafe code but also helps a fuzzer to find more unsafe code locations and thus execute the parts of a Rust program that are subject to memory safety issues.

Notably, in both evaluations, \toolname never performs worse compared to \aflrs when considering the performance over the 30 repetitions, regardless if we use the time it takes to trigger unsafe code or the number of unsafe code locations found. Hence, we argue that, on average, \toolname always outperforms \aflrs when testing unsafe code of Rust programs.

During our experiments \toolname is able to find one unique crash in \texttt{syn}, 12 unique crashes in \texttt{image}, one unique crash in \texttt{rust-cssparser}, and one unique crash in \texttt{toml}. To detect unique crashes, we did not rely on AFL's crash deduplication but instead wrote a custom script that analyzes stack traces returned by Rust's debug instrumentation. Further analyzed each set of crashes and if necessary, provided the resulting information to the responsible entities to fix the respective issue.

In the following we discuss the need for accurate call graph data and its influence on partial instrumentation and its performance when fuzzing a Rust program.

\begin{figure}[htb]
\center
\includegraphics[width=0.95\columnwidth]{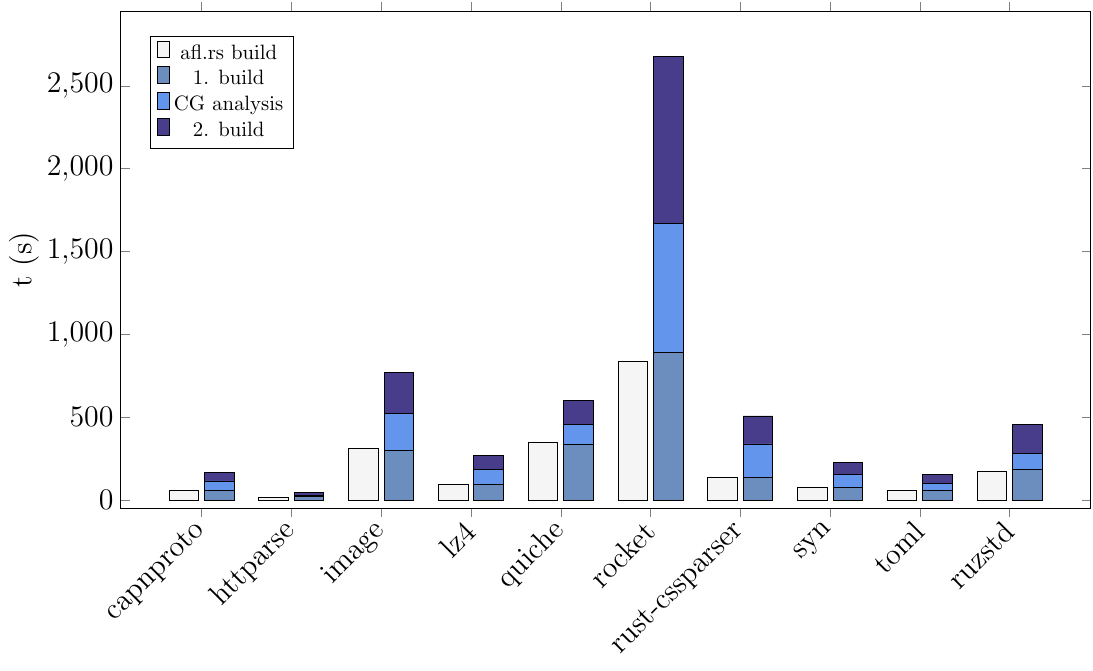}
\caption{Comparison of the compilation times of \aflrs and \toolname which needs to build the target twice and generate the list of functions that only execute safe Rust code to create a partially instrumented program.}
\label{fig:comp_compare}
\end{figure}

\subsubsection{Case Study: Naga}
\toolname relies on call graph data generated by LLVM to detect functions that do not require instrumentation. Thus, if the call graph data is erroneous, e.g., due to missing edges, the block list might contain errors as well. To better assess the importance of call graph data, we extended our analysis to programs outside of our test set. We use \emph{naga}, a popular shader translation library, with over 5 million downloads as a case study. Naga has a parser component that consumes tokens from the input data and calls corresponding handler functions, e.g., to generate the correct objects in memory. However, we observe that LLVM fails to generate the call edge between the front end parser and functions that are responsible to handle the parsed objects. As calling these handler functions leads to the execution of unsafe Rust code in the naga binary, they should be instrumented but due to the erroneous call graph data they are not. Note that initial tests show that missing this specific call edge does not prevent \toolname from executing the corresponding unsafe code locations but it requires more time compared to a fully instrumented fuzz target. Furthermore, we want to emphasize that this is not an inherent limitation of \toolname or partial instrumentation in general but a practical drawback due to the fact that LLVM cannot correctly determine all call targets at compile time.
Our case study highlights the need for proper call graph data when relying on it as part of partial code instrumentation during fuzzing.

\subsection{Compilation Time Overhead}

As mentioned before, the design of \toolname does \emph{not} impose any performance overhead when fuzzing a target as we do not require any additional instrumentation or complex analysis during run time. However, to generate the block list \toolname needs to create a call graph first, and subsequently compile the program again to add the code coverage feedback at the correct locations. Thus, we need to compile the program twice and additionally need to create a call graph and execute path reachability analysis using the \emph{path finder} component. We measure the time the compilation process takes for each program in our test set and provide the results in Figure~\ref{fig:comp_compare}. We find that the average performance overhead imposed by our implementation is around 6 minutes (175\%).
Given that programs are commonly fuzzed for several hours and up to multiple months, we argue that this overhead during the compilation process is reasonable and acceptable in practice.

\section{Related Work}
\label{sec:relatedwork}
In this section we discuss related work, i.e., other academic publications in areas related to \toolname and explain how our approach differs from them. Furthermore, we discuss other approaches that address the security of Rust libraries and programs.

\subsection{Fuzzing}

As we are not aware of any fuzzer that has been published in academia that focus on unsafe Rust code, the most closely related fuzzers are directed fuzzers for C and \Cpp code. The first directed fuzzer in academia has been implemented by \citet{BOE2017} and is called \emph{AFLGo}. In directed fuzzing, the idea is to approximate the distance of a target \emph{for every basic block} in a program and use this information as additional execution feedback. However, due to the fact that the distance calculation does not scale linearly with the number of basic blocks, this process can take a significant amount of time for complex real-world programs~\cite{HON2018} which can be a problem in practice, e.g., for usage in continuous integration frameworks that may employ short fuzz runs to test each commit to the code base. In contrast, partial instrumentation only requires call graph information which improves scaling for large code bases and does not require additional computations during run time.

Furthermore, AFLGo requires the implementation of a \emph{cooling schedule} which adjusts the probability that an input is assigned energy based on the distance to a target block, i.e., the importance of the distance calculation decreases over time. \toolname on the other hand, does not require a cooling schedule due to the usage of partial instrumentation, which works seamlessly with any of \aflpp's power schedules. 

Other directed fuzzers have similar issues as AFLGo (e.g., Hawkeye~\cite{HON2018}) or are not applicable to the challenges posed by Rust unsafe code fuzzing, e.g., AFLChurn~\cite{ZHU21} is specialized on patch testing, which is orthogonal to the challenge of testing unsafe Rust.

\textsc{CrabSandwich}~\cite{CRU2023} by \citeauthor{CRU2023} is a fuzzer that is specialized on fuzzing Rust programs and was published as a registered report including preliminary results.
\textsc{CrabSandwich} is a drop-in replacement for \texttt{cargo-fuzz} and is based on LibAFL~\cite{FIO2022}. At the time of writing, the source code of \textsc{CrabSandwich} is not available for testing.
\textsc{CrabSandwich} does not provide any means to specifically target unsafe code in a Rust program but rather treats all code coverage equal, similar to \aflrs. Thus, \toolname tackles a different set of challenges.

\subsection{Rust Security}

Researchers have analyzed the usage of unsafe code in publicly available Rust programs, i.e., crates from \url{crates.io}. \citet{AST2020} state that $21.3\%$ of all crates contain at least one explicit unsafe Rust statement, which the authors consider a significantly high number of crates. \citet{EVA2020} find that about $50\%$ of all crates utilize unsafe code, either directly or in dependency code. Furthermore, both studies analyzed how much of the code is actually written in unsafe Rust. \citeauthor{EVA2020} find that $90\%$ of projects that contain unsafe code use fewer than 10 unsafe code blocks, while \citeauthor{AST2020} state that most unsafe code blocks ($75\%$) consist of 21 or fewer MIR instructions. This shows that 
\begin{inparaenum}[(1)]
\item a significant number, but especially popular Rust projects, can profit from a fuzzer that focuses on unsafe code, and
\item that using partial instrumentation is useful due to the fact that unsafe code is used only for small parts of the program, and therefore most code is inherently memory safe.
\end{inparaenum}

Furthermore, researchers have also analyzed reported bugs found in real-world Rust programs. \citet{QIN2020} as well as \citet{XU2021} systematically analyzed security related Rust bugs and find that all of them involve unsafe Rust code. The only exception is a single bug in a pre-release Rust compiler (v0.3) which \citeauthor{QIN2020} consider insignificant. Both studies conclude that \emph{safe} Rust code can be considered safe based on the empirical data. This shows that focusing on unsafe code is a sound approach when testing Rust programs with automatic fuzz testing as implemented in \toolname.

Publications have also tackled the problem of finding bugs in Rust without utilizing dynamic analysis.
Rudra~\cite{BAE2021} as well as MirChecker~\cite{ZHU2021} are static analysis frameworks that utilize intermediate code representations to detect different bug types. While Rudra relies mostly on data-flow analysis to detect bugs, MirChecker utilizes a combination of numerical and symbolic analysis on the Rust MIR. Both implementations only support very specific bug types while fuzzers commonly support a wide variety of bug types. Furthermore, both tools suffer from a very high false positive rate of up to 80\% and 95\% respectively. Due to the fact that a fuzzer generates a concrete input that triggers a bug oracle, false positives are very unlikely as long as a proper fuzz target is used. During our experiments, \toolname did not cause any false positives. Furthermore, the generated crashing input allows debugging of the unwanted behavior and ultimately fixing the respective bug.

A different dynamic approach to find bugs in Rust libraries has been introduced by \citet{TAK2021} who implement SyRust. The idea is to utilize program synthesis to generate programs that call API functions of a Rust library. SyRust generates Rust programs which are compiled to MIR and executed via the Miri~\cite{SCH2016} interpreter which also acts as a bug oracle to detect unwanted behavior. Contrary to this, \toolname executes test inputs on native code and generates new inputs through different mutations that may cause program crashes which indicate a bug. This allows us to test a large number of test cases per second without any additional overhead.

Researchers have also put effort into improving other aspects of the fuzzing process. Namely, RULF~\cite{JIA2021}, RPG~\cite{XU2024}, and FRIES~\cite{XIZ2024} automatically generate fuzz targets which can be used to test as much code of a program under test as possible. Note that automatically generated fuzz targets can cause a large number of false positive crashes. For example, \citeauthor{JIA2021} report that fuzzers were able to find 636 unique crashes (i.e., after deduplication) but found only 30 actual bugs.

All three implementations generate API call sequences that process the fuzzer generated input. Thus, they address a different problem as they do not change the fuzzer itself but provide usable fuzz targets. \toolname improves the fuzzing process by utilizing partial instrumentation which prioritizes code locations that contain potentially memory unsafe Rust code. Furthermore, the existing fuzz target generation implementations are focused to test Rust library code whereas \toolname can be used to test library code as well as whole projects. Note that existing fuzzing recommendations state a fuzzing harness or target should be written and selected by an expert with domain knowledge~\cite{HAZ2020}.

Another component of the fuzzing process which has been the subject of scientific publications are bug oracles. \citet{MIN2024} as well as \citet{CHO2024} present Rust specific AddressSanitizer~(ASan) based optimizations. Both approaches try to minimize the number of necessary memory access checks during fuzz testing. While ERASan~\cite{MIN2024} considers raw pointers as the only source of potentially unwanted behavior, RustSan~\cite{CHO2024} includes any data object that is modified by unsafe Rust code. Additionally, RustSan implements more fine grained access controls to detect more memory access violations. However, due to the usage of computation intensive static analysis (e.g., points-to analysis) both approaches require a considerable compile-time overhead (of up to 31x) and are therefore not applicable to, e.g., integration into a CI pipeline. Note that both approaches may considerably reduce the number of existing memory access checks but this does not necessarily correlate to a similar performance improvement during fuzzing~\cite{CHO2024}. Furthermore, note that ASan is known to suffer from a considerable number of false negatives (i.e., inputs that crash a normal target may not crash the ASan instrumented program)~\cite{LI2021}.

Another line of work tries to isolate unsafe Rust code from safe Rust code (e.g., to separate Rust and C/\Cpp code). Code isolation approaches have been implemented in Sandcrust~\cite{LAM2017}, XRust~\cite{LIU2020}, Galeed~\cite{RIV2021}, or TRust~\cite{BAN2023}. Each approach protects different parts of the Rust program (e.g., heap data, foreign code, or untrusted objects) and utilizes different techniques to implement the separation of trusted and untrusted code (e.g., via guard pages, a separate process, or Intel MKP). Code isolation approaches always require additional security checks at run time which causes a significant performance overhead. Furthermore, isolation approaches do not fix the underlying issue, i.e., the bug or vulnerability in the program. \toolname identifies the security-related bugs and allows to fix them by providing a crash input that helps programmers to understand the corresponding code issue.

\section{Conclusion}
\label{sec:conclusion}
Our work shows that partial instrumentation for Rust fuzzing is a promising direction for security testing for Rust programs. 
Since the underlying design of safe Rust does successfully prevent memory corruption issues as well as data races, it is natural to prioritize unsafe Rust code when trying to detect security issues in Rust. Given that our approach does not require considerable manual effort but works automatically (i.e., selection of irrelevant functions as well as instrumentation) it is as simple to use as \aflrs. Our evaluation shows that partial instrumentation can be a powerful technique to improve the performance of unsafe Rust code fuzzing. \toolname requires statistically significantly less time when generating inputs that execute unsafe code blocks and trigger more unsafe code locations compared to \aflrs. Our paper shows for the first time that partial instrumentation is a viable path to improve a fuzzing implementation for Rust code without imposing any overhead during the fuzz testing process.

\subsection*{Acknowledgment}
This work has been funded by the Deutsche Forschungsgemeinschaft (DFG, German Research Foundation) under Germany's Excellence Strategy - EXC 2092 CASA - 390781972, and CROSSING (SFB 1119) 236615297 within project~T1.

\bibliographystyle{plainnat}
{\footnotesize
\bibliography{references}}

\end{document}